\documentclass[conference,final,twocolumn,10pt]{IEEEtran} % for single column no footer

\usepackage{cite}
\ifCLASSINFOpdf
  \usepackage[pdftex]{graphicx}
  \usepackage{epstopdf} % remove this
\else
\fi
\usepackage{amsmath}
\usepackage{algorithm}
\usepackage{algorithmic}

  \usepackage[caption=false,font=footnotesize]{subfig}
\usepackage{url}
\usepackage{amsthm}

\theoremstyle{definition}

\theoremstyle{remark}

\newcommand{\comment}[1]{{}}

\usepackage{amssymb}
\usepackage{xspace}
\usepackage{bm}

\newcommand{\set}[1]{\ensuremath{\mathcal{#1}}\xspace} % caligraphic set notation
\newcommand{\mat}[1]{\ensuremath{\mathbf{#1}}\xspace} % matrices
\renewcommand{\vec}[1]{\ensuremath{\mathbf{#1}}\xspace} % vectors
\newcommand{\parens}[1]{\ensuremath{\left(#1\right)}\xspace}
\newcommand{\brackets}[1]{\ensuremath{\left[#1\right]}\xspace}
\newcommand{\braces}[1]{\ensuremath{\left\{#1\right\}}\xspace}
\newcommand{\bars}[1]{\ensuremath{\left\vert#1\right\vert}\xspace}
\newcommand{\doublebars}[1]{\ensuremath{\left\Vert#1\right\Vert}\xspace}

\newcommand{\complex}{\ensuremath{\mathbb{C}}\xspace}

\renewcommand{\j}{\ensuremath{\mathrm{j}}}

\newcommand{\card}[1]{\bars{#1}}

\newcommand{\setcomplex}{\ensuremath{\complex}}

\newcommand{\setvector}[2]{\ensuremath{#1^{#2 \times 1}}\xspace}

\newcommand{\setvectorcomplex}[1]{\setvector{\setcomplex}{#1}}

\newcommand{\setmatrix}[3]{\ensuremath{#1^{#2 \times #3}}\xspace}

\newcommand{\setmatrixcomplex}[2]{\setmatrix{\setcomplex}{#1}{#2}}

\newcommand{\onevec}[1]{\ensuremath{\vec{1}_{#1}}\xspace}
\newcommand{\zeromat}{\ensuremath{\mat{0}}\xspace}

\newcommand{\inv}{\ensuremath{^{-1}}\xspace}

\newcommand{\ctrans}{\ensuremath{^{{*}}}\xspace}
\newcommand{\entry}[2]{\ensuremath{\brackets{#1}_{#2}}\xspace}

\newcommand{\logtwo}[1]{\ensuremath{\mathrm{log}_{2}\parens{#1}}}

\newcommand{\diag}[1]{\ensuremath{\mathrm{diag}\parens{#1}}}

\newcommand{\pnorm}[2]{\ensuremath{\doublebars{#2}_{#1}}\xspace}

\newcommand{\normtwo}[1]{\pnorm{2}{#1}}

\newcommand{\normfro}[1]{\pnorm{\mathrm{F}}{#1}}

\newcommand{\ev}[1]{\ensuremath{\mathbb{E}\brackets{#1}}\xspace}

\newcommand{\minop}[1]{\ensuremath{\mathrm{min}\parens{#1}}\xspace}

\newcommand{\st}{\ensuremath{\mathrm{s.t.~}}\xspace}
\newcommand{\opt}{\ensuremath{^{\star}}\xspace}
\newcommand{\project}[2]{\ensuremath{\Pi_{#2}\parens{#1}}\xspace}

\newcommand{\snr}{\ensuremath{\mathrm{SNR}}\xspace}

\newcommand{\inr}{\ensuremath{\mathrm{INR}}\xspace}

\newcommand{\atx}[1]{\ensuremath{\vec{a}_{\mathrm{tx}}\parens{#1}}\xspace}
\newcommand{\arx}[1]{\ensuremath{\vec{a}_{\mathrm{rx}}\parens{#1}}\xspace}

\newcommand{\Atx}{\ensuremath{\mA_{\mathrm{tx}}}\xspace}
\newcommand{\Arx}{\ensuremath{\mA_{\mathrm{rx}}}\xspace}

\newcommand{\Nt}{\ensuremath{N_\mathrm{t}}\xspace} % number of antennas at Tx
\newcommand{\Nr}{\ensuremath{N_\mathrm{r}}\xspace} % number of antennas at Rx
\newcommand{\precbmat}{\ensuremath{{\mat{F}}}\xspace}
\newcommand{\comcbmat}{\ensuremath{{\mat{W}}}\xspace}

\newcommand{\precb}{\ensuremath{\mathcal{F}}\xspace}
\newcommand{\precbbar}{\ensuremath{\bar{\mathcal{F}}}\xspace}

\newcommand{\comcb}{\ensuremath{\mathcal{W}}\xspace}
\newcommand{\comcbbar}{\ensuremath{\bar{\mathcal{W}}}\xspace}

\newcommand{\Gtx}{\ensuremath{G_{\mathrm{tx}}}\xspace}
\newcommand{\Grx}{\ensuremath{G_{\mathrm{rx}}}\xspace}

\newcommand{\Gtxmax}{\ensuremath{G_{\mathrm{tx,max}}}\xspace}
\newcommand{\Grxmax}{\ensuremath{G_{\mathrm{rx,max}}}\xspace}
\newcommand{\Gtxtgt}{\ensuremath{G_{\mathrm{tx,tgt}}}\xspace}
\newcommand{\Grxtgt}{\ensuremath{G_{\mathrm{rx,tgt}}}\xspace}

\newcommand{\gtx}{\ensuremath{\vg_{\mathrm{tx}}}\xspace}
\newcommand{\grx}{\ensuremath{\vg_{\mathrm{rx}}}\xspace}

\newcommand{\gtxtgt}{\ensuremath{\gtx^{\mathrm{tgt}}}\xspace}
\newcommand{\grxtgt}{\ensuremath{\grx^{\mathrm{tgt}}}\xspace}

\newcommand{\Deltatx}{\ensuremath{\Delta_{\mathrm{tx}}}\xspace}
\newcommand{\Deltarx}{\ensuremath{\Delta_{\mathrm{rx}}}\xspace}

\newcommand{\sigmatx}{\ensuremath{\sigma_{\mathrm{tx}}}\xspace}
\newcommand{\sigmarx}{\ensuremath{\sigma_{\mathrm{rx}}}\xspace}

\newcommand{\bitsamp}{\ensuremath{{b_{\mathrm{amp}}}}\xspace}
\newcommand{\bitsphase}{\ensuremath{{b_{\mathrm{phs}}}}\xspace}

\newcommand{\Hsiset}{\ensuremath{\set{H}}}

\newcommand{\capfd}{\ensuremath{C_{\mathrm{fd}}}\xspace}
\newcommand{\caphd}{\ensuremath{C_{\mathrm{hd}}}\xspace}

\newcommand{\setx}{\ensuremath{R_{\mathrm{tx}}}\xspace}
\newcommand{\serx}{\ensuremath{R_{\mathrm{rx}}}\xspace}

\newcommand{\vhtx}{\ensuremath{\vh_{\mathrm{tx}}}\xspace}
\newcommand{\vhrx}{\ensuremath{\vh_{\mathrm{rx}}}\xspace}

\newcommand{\snrtx}{\ensuremath{\snr_{\mathrm{tx}}}\xspace}
\newcommand{\snrrx}{\ensuremath{\snr_{\mathrm{rx}}}\xspace}

\newcommand{\idx}[1]{\ensuremath{^{\parens{#1}}}\xspace}

\newcommand{\dir}{\ensuremath{\vartheta}\xspace}
\newcommand{\dirtx}{\ensuremath{\dir_{\mathrm{tx}}}\xspace}
\newcommand{\dirrx}{\ensuremath{\dir_{\mathrm{rx}}}\xspace}

\newcommand{\dirtxset}{\ensuremath{\mathcal{A}_{\mathrm{tx}}}\xspace}
\newcommand{\dirrxset}{\ensuremath{\mathcal{A}_{\mathrm{rx}}}\xspace}

\newcommand{\Mtx}{\ensuremath{{M}_{\mathrm{tx}}}\xspace}
\newcommand{\Mrx}{\ensuremath{{M}_{\mathrm{rx}}}\xspace}

\def\vf{{\vec{f}}}
\def\vg{{\vec{g}}}
\def\vh{{\vec{h}}}

\def\vv{{\vec{v}}}
\def\vw{{\vec{w}}}

\def\mA{{\mat{A}}}

\def\mH{{\mat{H}}}

\def\mDelta{{\mat{\Delta}}}

\usepackage{xspace}
\usepackage[acronym,nogroupskip,nonumberlist,nopostdot]{glossaries}
\makeglossaries

\usepackage{xcolor}

\newacronym{snr}{SNR}{signal-to-noise ratio}
\newacronym{sinr}{SINR}{signal-to-interference-plus-noise ratio}
\newacronym{inr}{INR}{interference-to-noise ratio}
\newacronym{sir}{SIR}{signal-to-interference ratio}
\newacronym{sqr}{SQR}{signal-to-quantization-noise ratio}
\newacronym{sqnr}{SQNR}{signal-to-quantization-plus-noise ratio}
\newacronym{ian}{IAN}{interference as noise}
\newacronym{ber}{BER}{bit error rate}
\newacronym{pn}{PN}{pseudorandom noise}
\newacronym{bfsk}{BFSK}{binary frequency shift keying}
\newacronym{fh}{FH}{frequency-hopped}
\newacronym{fh-bfsk}{FH-BFSK}{frequency-hopped binary frequency shift keying}
\newacronym{crc}{CRC}{cyclic redundancy check}
\newacronym{isi}{ISI}{intersymbol interference}
\newacronym{dsss}{DSSS}{direct-sequence spread spectrum}
\newacronym{ofdm}{OFDM}{orthogonal frequency-division multiplexing}
\newacronym{ofdma}{OFDMA}{orthogonal frequency-division multiple access}
\newacronym{sdr}{SDR}{software-defined radio}
\newacronym{tx}{TX}{transmitter}
\newacronym{rx}{RX}{receiver}
\newacronym{fdd}{FDD}{frequency-division duplexing}
\newacronym{tdd}{TDD}{time-division duplexing}
\newacronym{fdma}{FDMA}{frequency-division multiple access}
\newacronym{tdma}{TDMA}{time-division multiple access}
\newacronym{sdma}{SDMA}{space-division multiple access}
\newacronym[plural=MPCs]{mpc}{MPC}{multipath component}
\newacronym{mui}{MUI}{multi-user interference}

\newacronym{qam}{QAM}{quadrature amplitude modulation}
\newacronym{mqam}{MQAM}{M-ary quadrature amplitude modulation}

\newacronym{ls}{LS}{least-squares}
\newacronym{lms}{LMS}{least mean squares}
\newacronym{rls}{RLS}{recursive least-squares}
\newacronym{rzf}{RZF}{regularized zero-forcing}
\newacronym{mmse}{MMSE}{minimum mean square error}
\newacronym{lmmse}{LMMSE}{linear minimum mean square error}
\newacronym{mse}{MSE}{mean square error}
\newacronym{fft}{FFT}{fast Fourier transform}
\newacronym{dft}{DFT}{discrete Fourier transform}
\newacronym{dtft}{DTFT}{discrete-time Fourier transform}
\newacronym{ctft}{CTFT}{continuous-time Fourier transform}
\newacronym{ml}{ML}{machine learning}
\newacronym[plural=NNs]{nn}{NN}{neural network}
\newacronym[plural=RNNs]{rnn}{RNN}{recurrent neural network}
\newacronym[plural=ADCs]{adc}{ADC}{analog-to-digital converter}
\newacronym[plural=DACs]{dac}{DAC}{digital-to-analog converter}
\newacronym[plural=FPGAs]{fpga}{FPGA}{field-programmable gate array}
\newacronym{evm}{EVM}{error vector magnitude}
\newacronym{enob}{ENOB}{effective number of bits}
\newacronym{zf}{ZF}{zero-forcing}
\newacronym{rv}{r.v.}{random variable}
\newacronym{omp}{OMP}{orthogonal matching pursuit}
\newacronym{svd}{SVD}{singular value decomposition}
\newacronym{lsb}{LSB}{least significant bit}

\newacronym{sdp}{SDP}{semidefinite programming}
\newacronym{psd}{PSD}{positive semidefinite}
\newacronym{nsd}{NSD}{negative semidefinite}

\newacronym{agc}{AGC}{automatic gain control}
\newacronym{rf}{RF}{radio frequency}
\newacronym{los}{LOS}{line-of-sight}
\newacronym{nlos}{NLOS}{non-line-of-sight}
\newacronym{ple}{PLE}{path loss exponent}
\newacronym[plural=dB,firstplural=decibels (dB)]{db}{dB}{decibel}
\newacronym[plural=dBm,firstplural=decibel milliwatts (dBm)]{dbm}{dBm}{decibel milliwatts}
\newacronym{pa}{PA}{power amplifier}
\newacronym{lna}{LNA}{low noise amplifier}
\newacronym{cw}{CW}{continuous wave}
\newacronym{papr}{PAPR}{peak-to-average power ratio}
\newacronym{usrp}{USRP}{Universal Software Radio Peripheral}
\newacronym{irr}{IRR}{image rejection ratio}
\newacronym{lo}{LO}{local oscillator}
\newacronym{vm}{VM}{vector modulator}
\newacronym{mmwave}{mmWave}{millimeter wave}
\newacronym{eirp}{EIRP}{effective isotropic radiated power}

\newacronym{csma}{CSMA}{carrier-sense multiple access}
\newacronym{csmaca}{CSMA/CA}{carrier-sense multiple access with collision avoidance}
\newacronym{csmacd}{CSMA/CD}{carrier-sense multiple access with collision detection}
\newacronym{mac}{MAC}{medium access control}
\newacronym{phy}{PHY}{physical layer}
\newacronym{4g}{4G}{fourth generation}
\newacronym{lte}{LTE}{Long-Term Evolution}
\newacronym{4glte}{4G LTE}{\gls{4g} \gls{lte}}
\newacronym{5g}{5G}{fifth generation}
\newacronym{nr}{NR}{New Radio}
\newacronym{5gnr}{5G NR}{5G New Radio}
\newacronym{ieee}{IEEE}{Institute of Electrical and Electronics Engineers}
\newacronym{wifi}{Wi-Fi}{IEEE 802.11}
\newacronym{lan}{LAN}{local area network}
\newacronym{wlan}{WLAN}{wireless local area network}
\newacronym[plural=BSs]{bs}{BS}{base station}
\newacronym[plural=SBSs]{sbs}{SBS}{small-cell base station}
\newacronym[plural=FD-SBSs]{fdsbs}{FD-SBS}{\gls{fd}-enabled \gls{sbs}}
\newacronym[plural=MBSs]{mbs}{MBS}{macrocell base station}
\newacronym[plural=UEs]{ue}{UE}{user equipment}
\newacronym{ul}{UL}{uplink}
\newacronym{dl}{DL}{downlink}
\newacronym{qos}{QoS}{Quality of Service}
\newacronym{fcc}{FCC}{Federal Communications Commission}
\newacronym{iab}{IAB}{integrated access and backhaul}
\newacronym{fab}{FAB}{fixed access and backhaul}
\newacronym{hetnet}{HetNet}{heterogeneous network}

\newacronym{siso}{SISO}{single-input single-output}
\newacronym{mimo}{MIMO}{multiple-input multiple-output}
\newacronym{sumimo}{SU-MIMO}{single-user \gls{mimo}}
\newacronym{mumimo}{MU-MIMO}{multi-user \gls{mimo}}
\newacronym{bf}{BF}{beamforming}
\newacronym{ca}{CA}{constant amplitude}
\newacronym{ula}{ULA}{uniform linear array}
\newacronym{upa}{UPA}{uniform planar array}
\newacronym[\glslongpluralkey={angles of arrival}]{aoa}{AoA}{angle of arrival}
\newacronym[\glslongpluralkey={angles of departure}]{aod}{AoD}{angle of departure}
\newacronym{dof}{DoF}{degrees of freedom}
\newacronym{csi}{CSI}{channel state information}
\newacronym{csit}{CSIT}{\gls{csi} at the transmitter}
\newacronym{csir}{CSIR}{\gls{csi} at the receiver}
\newacronym{cs}{CS}{compressed sensing}

\newacronym{fd}{FD}{in-band full-duplex}
\newacronym{hd}{HD}{half-duplex}
\newacronym{si}{SI}{self-interference}
\newacronym{sic}{SIC}{self-interference cancellation}
\newacronym{soi}{SoI}{signal of interest}
\newacronym{asic}{A-SIC}{analog \acrlong{sic}}
\newacronym{dsic}{D-SIC}{digital \gls{sic}}
\newacronym{star}{STAR}{simultaneous transmit and receive}
\newacronym{warp}{WARP}{Wireless Open-Access Research Platform}
\newacronym{bfc}{BFC}{beamforming cancellation}
\newacronym{ipi}{IPI}{inter-panel-interference}
\newacronym{ipic}{IPIC}{inter-panel-interference cancellation}

\newacronym{qcqp}{QCQP}{quadratically-constrained quadratic programming}
\newacronym{cdf}{CDF}{cumulative density function}

\newacronym{elf}{ELF}{extremely low frequency}
\newacronym{slf}{SLF}{super low frequency}
\newacronym{ulf}{ULF}{ultra low frequency}
\newacronym{vlf}{VLF}{very low frequency}
\newacronym{lf}{LF}{low frequency}
\newacronym{mf}{MF}{medium frequency}
\newacronym{hf}{HF}{high frequency}
\newacronym{vhf}{VHF}{very high frequency}
\newacronym{uhf}{UHF}{ultra high frequency}
\newacronym{shf}{SHF}{super high frequency}
\newacronym{ehf}{EHF}{extremely high frequency}
\newacronym{thf}{THF}{tremendously high frequency}

\newacronym{wncg}{WNCG}{Wireless Networking and Communications Group}
\newacronym{linc}{LINC}{Laboratory of Informatics, Networks, and Communications}
\newacronym{ut}{UT Austin}{The University of Texas at Austin}
\newacronym{uiuc}{UIUC}{University of Illinois at Urbana-Champaign}
\newacronym{usc}{USC}{University of Southern California}
\newacronym{mit}{MIT}{Massachusetts Institute of Technology}
\newacronym{berkeley}{UC Berkeley}{University of California, Berkeley}
\newacronym{osu}{OSU}{Ohio State University}

\newcommand{\mmwave}{\gls{mmwave}\xspace}

\newcommand{\iab}{\gls{iab}\xspace}

\newcommand{\upas}{\glspl{upa}\xspace}

\newcommand{\figref}[1]{\figurename~\ref{#1}}

\begin{document}
	
% \title{Analog Beamforming Codebook Design for Millimeter Wave Full-Duplex}
% \title{Analog Beamforming Codebook Design for Millimeter Wave Full-Duplex}
% \title{Millimeter Wave Analog Beamforming\\Codebooks Robust to Self-Interference}
\title{Millimeter Wave Analog Beamforming\\Codebooks Robust to Self-Interference}
% \title{\textsc{Steer}: Millimeter Wave Analog Beamforming Codebooks Robust to Self-Interference}
% \title{Analog Beamforming Codebook Design for Millimeter-Wave Interference Channels}

\author{
    \IEEEauthorblockN{%
        Ian~P.~Roberts\IEEEauthorrefmark{1}, %
        Hardik~B.~Jain\IEEEauthorrefmark{2}, %
        Sriram~Vishwanath\IEEEauthorrefmark{2}, %
        and Jeffrey~G.~Andrews\IEEEauthorrefmark{1} %
    }
    \IEEEauthorblockA{\IEEEauthorrefmark{1}University of Texas at Austin, Austin, TX, USA.}
    \IEEEauthorblockA{\IEEEauthorrefmark{2}GenXComm,~Inc., Austin, TX, USA.}
}

\maketitle

% \pagebreak

% \setcounter{tocdepth}{1} % 0,1,2,...
% \tableofcontents

% \pagebreak

\begin{abstract}
% Abstract here.

This paper develops a novel methodology for designing analog beamforming codebooks for full-duplex \mmwave transceivers, the first such codebooks to the best of our knowledge.
% Our codebooks are designed to facilitate full-duplexing transmission and reception by maintaining high beamforming gain while reducing self-interference coupled between transmit and receive beams of a full-duplex transceiver.
Our design reduces the self-interference coupled by transmit-receive beam pairs and simultaneously delivers high  beamforming gain over desired coverage regions, allowing \mmwave full-duplex systems to support beam alignment while minimizing self-interference. 
To do so, our methodology allows some variability in beamforming gain to strategically shape beams that reject self-interference while still having substantial gain.
We present an algorithm for approximately solving our codebook design problem while accounting for the non-convexity posed by digitally-controlled phase shifters and attenuators.
Numerical results suggest that our design can outperform or nearly match existing codebooks in sum spectral efficiency across a wide range of self-interference power levels.
Results show that our design offers an extra 20--50 dB of robustness to self-interference, depending on hardware constraints.
\end{abstract}

% \input{sec-abstract.tex}

% \pagebreak

% \input{sec-peer-review-title.tex}

% \input{sec-keywords.tex}

\glsresetall

% \pagebreak

\section{Introduction} \label{sec:introduction}

Codebook-based analog beamforming is a critical component of modern \mmwave systems \cite{heath_overview_2016}.
Rather than measure the complete over-the-air channel and subsequently configure analog beamformers, current \mmwave systems instead rely on beam alignment procedures to more quickly identify promising transmit and receive beamforming directions, typically via exploration of a codebook of candidate analog beams \cite{heath_overview_2016,junyi_wang_beam_2009}.
This offers a simple and robust way to configure dozens of phase shifters and attenuators without extensive channel knowledge \textit{a priori}.
% This offers a reduction in complexity versus setting dozens of phase shifters and attenuators one-by-one.

Designing analog beamforming codebooks that span a desired coverage region is relatively straightforward for conventional half-duplex \mmwave systems \cite{junyi_wang_beam_2009}.
% Depending on the phase shifter and attenuator resolution, number of antennas, and desired steering resolution, a set of highly directional beams that span a desired coverage region can be fairly easily designed.
% A \gls{dft} codebook, for example, is a simple set of beams that serve desired directions and only requires phase shifters, not attenuators \cite{heath_lozano}.
Conjugate beamforming (i.e., matched filter beamforming), for example, is a simple way to construct a set of beams that serve desired directions and only requires phase shifters, not attenuators \cite{heath_lozano}.
% With sufficient phase resolution, conjugate beams can provide full array gain in any desired direction.
Other designs leverage attenuators to shape beams that exhibit wider main lobes and suppress side lobes, for example, which can reduce beam misalignment losses and inter-user interference.

Equipping \mmwave systems with full-duplex capability is an attractive proposition at not only the physical layer but also as a deployment solution through \iab \cite{roberts_wcm}.
To enable \mmwave full-duplex, recent work \cite{satyanarayana_hybrid_2019,liu_beamforming_2016,lopez_prelcic_2019_analog,lopez-valcarce_beamformer_2019,prelcic_2019_hybrid,zhu_uav_joint_2020,cai_robust_2019,roberts_2019_bfc,roberts_equipping_2020,roberts_bflrdr} has explored beamforming-based self-interference mitigation in various contexts.
Some designs \cite{liu_beamforming_2016,satyanarayana_hybrid_2019,prelcic_2019_hybrid,lopez_prelcic_2019_analog,lopez-valcarce_beamformer_2019,zhu_uav_joint_2020,cai_robust_2019} have neglected codebook-based analog beamforming, assuming the ability to fine-tune each phase shifter dynamically, and do not account for \textit{digitally}-controlled phase shifters.
Most designs have assumed a lack of amplitude control, even though it is not uncommon to have both phase shifters and attenuators in practical analog beamforming networks.
In addition, existing solutions \cite{liu_beamforming_2016,satyanarayana_hybrid_2019,prelcic_2019_hybrid,lopez_prelcic_2019_analog,lopez-valcarce_beamformer_2019,zhu_uav_joint_2020,cai_robust_2019,roberts_2019_bfc} have assumed transmit and receive channel knowledge along with that of the self-interference channel to configure precoding and combining---meaning they neglect beam alignment and rely on highly dynamic updates as the transmit and receive channels change.

Like half-duplex \mmwave systems, one with full-duplex capability will presumably conduct codebook-based beam alignment on its transmit link \textit{and} receive link, meaning it will juggle a transmit beam and receive beam concurrently, which couple together via the self-interference channel.
Off-the-shelf analog beamforming codebooks that were designed for half-duplex settings may be undesirable in full-duplex settings since they do not necessarily offer robustness to self-interference \cite{roberts_wcm}.
% This is because, at a full-duplex \mmwave transceiver, the transmit and receive beams couple together via the self-interference channel. 
% Off-the-shelf analog beamforming codebooks used for half-duplex, do not necessarily offer robustness to rejecting this self-interference---after all, they were not designed for such.
Instead, we design analog beamforming codebooks for full-duplex that reliably deliver high beamforming gain to users and simultaneously reject self-interference regardless of which transmit and receive beams are used.  Given such a codebook, standard beam alignment procedures that are self-interference agnostic can be utilized in a full-duplex system.
This is a desirable practical outcome from our approach. 
% Moreover, we aim to achieve such regardless of which transmit and receive beams are used from their respective codebooks.
% By doing so, improved isolation between the transmitter and receiver of a full-duplex device can improve the degree of self-interference mitigation afforded by analog beamforming.

Among existing literature, we are not aware of any work on the design of analog beamforming \textit{codebooks} for \mmwave full-duplex.  Our main contribution is a methodology for designing transmit and receive analog beamforming codebooks that reduces the average self-interference coupled between transmit-receive beam pairs while also guaranteeing the beamforming gain they provide. We present an algorithm for approximately solving for our design that addresses the non-convexity posed by digitally-controlled phase shifters and attenuators.
Results indicate that our design offers $20$--$50$ dB of added robustness to self-interference by strategically shaping beams with gain comparable to conventional codebooks. 

As a result, codebooks designed with our approach could improve existing \mmwave full-duplex work \cite{roberts_2019_bfc,cai_robust_2019,roberts_equipping_2020,roberts_bflrdr} that accounts for codebook-based analog beamforming.  Hybrid digital/analog beamforming full-duplex systems could leverage our codebooks by weakening the \textit{effective} self-interference channel post beam alignment. Perhaps most excitingly, a \mmwave system employing our codebook design can in principle execute beam alignment and then seamlessly operate in a full-duplex fashion, thanks to codebooks whose beams offer inherent robustness to self-interference.  % meaning additional measures to mitigate such would not be necessary.

\section{System Model} \label{sec:system-model}

\begin{figure}
    \centering
    \includegraphics[width=\linewidth,height=\textheight,keepaspectratio]{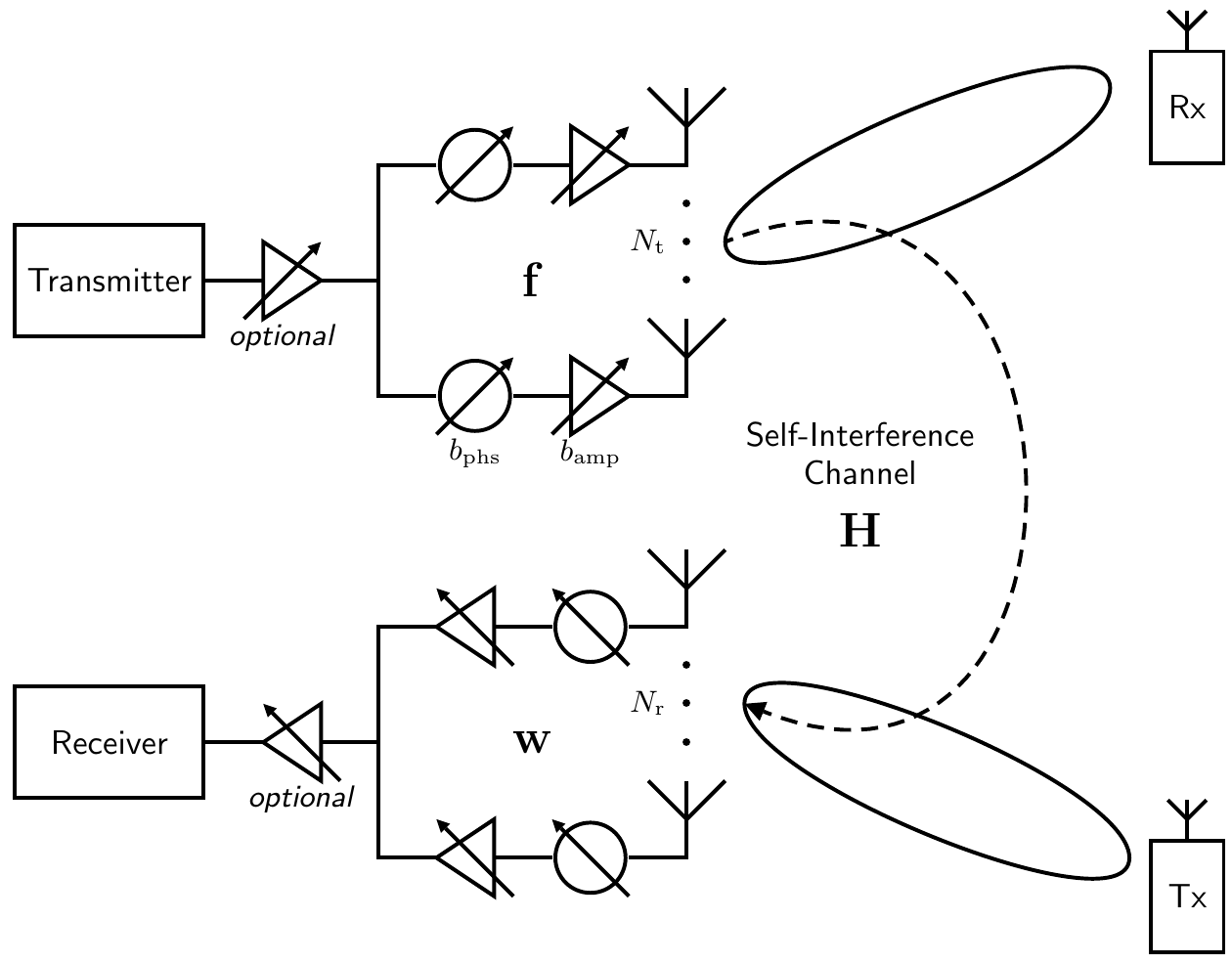}
    % \caption{A full-duplex \mmwave transceiver with separate transmit and receive arrays.}
    \caption{A full-duplex \mmwave transceiver with analog beamforming serving an uplink user with its receive beam and downlink user with its transmit beam.}
    \label{fig:system}
\end{figure}

In this work, we consider an in-band full-duplex \mmwave transceiver that employs separate, independently controlled arrays for transmission and reception, as illustrated in \figref{fig:system}.
Let $\Nt$ and $\Nr$ be the number of transmit and receive antennas, respectively, at the arrays of the full-duplex transceiver.
We denote as $\atx{\dir} \in \setvectorcomplex{\Nt}$ and $\arx{\dir} \in \setvectorcomplex{\Nr}$ the transmit and receive array response vectors, respectively, in the direction $\dir$.
% These array response vectors for arbitrary $\dir$ can be computed based on the transmit and receive array geometries.
We use the convention where $\normtwo{\atx{\dir}}^2 = \Nt$ and $\normtwo{\arx{\dir}}^2 = \Nr$.

We consider an analog-only beamforming system, though the codebook design herein could also be used by hybrid beamforming systems.
Let $\vf \in \setvectorcomplex{\Nt}$ be the analog precoding vector used at the transmitter and $\vw \in \setvectorcomplex{\Nr}$ be the analog combining vector used at the receiver.
% Analog beamforming is implemented in hardware where a network of phase shifters and attenuators are used to realize a desired beamformer.
We consider the practical case where the phase shifters and attenuators of the analog beamforming networks are digitally-controlled with $\bitsphase$ and $\bitsamp$ bits of resolution, respectively.
% Let $\bitsphase$ and $\bitsamp$ be the number of bits used to configure each phase shifter and attenuator in the transmit and receive beamforming networks, respectively.
We assume the set of all $2^\bitsphase$ possible phase shifter settings are uniformly distributed over the range $[0,2\pi)$ radians. 
% Since transmit and receive beamforming networks are power-constrained \textit{per-antenna} (rather than enjoying an $\ell_2$-based \textit{total} power constraint), 
We enforce that each beamforming weight not exceed unit magnitude, 
%\begin{gather}
%\bars{\entry{\vf}{m}} \leq 1 \ \forall \ m = 1, \dots, \Nt \\
%\bars{\entry{\vw}{n}} \leq 1 \ \forall \ n = 1, \dots, \Nr
%\end{gather}
understood by the fact that attenuators are used to implement amplitude control.
% Consequently, we can see that $\normtwo{\vf}^2 \leq \Nt$ and $\normtwo{\vw}^2 \leq \Nr$.
We assume the set of all $2^\bitsamp$ possible attenuator settings are (arbitrarily) distributed over the range $(0,1]$; 
% For example, digitally-stepped attenuators used in phased arrays may
it is practical to have an attenuation step size of $0.25$ or $0.5$ dB per \gls{lsb}.
% When amplitude control is not present, we can let $\bitsamp = 0$.
Capturing limited phase and amplitude control, let $\precb \subset \setvectorcomplex{\Nt}$ and $\comcb \subset \setvectorcomplex{\Nr}$ be the sets of all possible beamforming vectors $\vf$ and $\vw$, respectively. % ; hence, $\vf \in \precb$ and $\vw \in \comcb$.
% Note that the sets $\precb$ and $\comcb$ are equivalent when $\Nt = \Nr$ since we have assumed the same phase shifters and attenuators are used at both.

Since transmission and reception happen simultaneously and in-band, a self-interference channel $\mH \in \setmatrixcomplex{\Nr}{\Nt}$ manifests between the transmit and receive arrays of the full-duplex transceiver.
It is difficult to reliably assume a characterization or model for $\mH$ given a lack of measurements.
% One may expect near-field coupling to dominate due to the close proximity of the arrays and the fact that reflections from the environment will undergo severe path loss at \mmwave.
% At present, there has been little research investigating \mmwave self-interference channels, though sensible speculations have been made that assume it to be comprised of a strong direct coupling between the arrays along with reflections off of the environment \cite{li_2014}.
% This has, in fact, been the basis for a number of existing works on \mmwave full-duplex.
% The highly idealized near-field model used therein, however, will likely not be present in practice due to the backplates, casings, enclosures, and infrastructure associated with actual arrays.
% Still, it is difficult to speculate what the channel will be comprised of since 
% The close proximity between the transmit array and receive array make it difficult to characterize $\mH$.
% At present, there has been little research investigating such a channel, though sensible speculations have been made that assume it to be comprised of a direct coupling between the arrays along with reflections off of the environment.
% Given that this model has not yet been verified with measurements, 
As such, we do not present a design based on any assumption of $\mH$ but will evaluate our design using a commonly used near-field model.
% We do, however, use the aforementioned channel model and others for $\mH$ to assess our design in \secref{sec:simulation-results} to evaluate how our approach generalizes across different types of self-interference channels.

% Let us assume that the channel matrix $\Hsi$ remains stationary across service to many users.

A transmitted symbol $s$ from the full-duplex transceiver will be injected into the self-interference channel $\mH$ by its transmit beam $\vf$ and transmit power gain $P$, then subsequently combined by its receive beam $\vw$.
To abstract out the large-scale gain of the self-interference channel from its spatial characteristics, we impose $\ev{\normfro{\mH}^2} = \Nt \cdot \Nr$
%\begin{align}
%\ev{\normfro{\mH}^2} = \Nt \cdot \Nr
%\end{align}
and let a scalar $G \geq 0$ handle proper scaling of $\mH$ based on the isolation between the arrays (practically, $G \ll 1$).
The received self-interference $y$ can thus be written as
\begin{align}
y = \sqrt{P} \cdot G \cdot \vw\ctrans \mH \vf s .
\end{align}
It can be readily seen that decreasing the large-scale gain $G$ (increasing isolation) would reduce self-interference as well as strategically steering $\vf$ and $\vw$ according to the channel $\mH$.

% \input{sec-system-model-v2.tex}

% \input{sec-system-model.tex}

% \pagebreak

% \input{sec-problem-formulation-v2.tex}

% \input{sec-problem-formulation.tex}

% \pagebreak

% \input{sec-contribution.tex}

% \pagebreak

\section{Analog Beamforming Codebook Design} \label{sec:contribution-i}

% Analog beamforming codebooks have become critical to successful deployment of \mmwave communication systems.
% Rather than create beams arbitrarily throughout deployment, practical systems rely on relatively small sets of beams, from which a beam can be selected to serve a specific user.
While transmit beams $\vf$ are \textit{fundamentally} limited to come from $\precb$ and likewise $\vw$ from $\comcb$, engineers establish smaller codebooks $\precbbar \subset \precb$ and $\comcbbar \subset \comcb$ from which $\vf$ and $\vw$ will be drawn, respectively (e.g., via beam alignment).
We refer to the sizes of these codebooks as $\card{\precbbar} = \Mtx$ and $\card{\comcbbar} = \Mrx$, which are much smaller than their counterparts $\precb$ and $\comcb$. % given that
%\begin{gather}
%\card{\precbbar} = \Mtx \ll \card{\precb} = 2^{\Nt\cdot\parens{\bitsphase + \bitsamp}} \\
%\card{\comcbbar} = \Mrx \ll \card{\comcb} = 2^{\Nr\cdot\parens{\bitsphase + \bitsamp}}
%\end{gather}
% The beams comprising codebooks \precbbar and \comcbbar should collectively supply high beamforming gain to combat severe \mmwave path loss and achieve sufficient link margin.
% \footnote{In hierarchical codebooks, which are often used for initial access and beam alignment, $\precbbar$ and $\comcbbar$ refer to the finest tier of beams that are used for service.}
% As such, a codebook of beams must provide adequate coverage over some service region to ensure that the location of a user within that region does not prevent it from being served with high beamforming gain.

A transmit beam $\vf$ and receive beam $\vw$, chosen from some codebooks $\precbbar$ and $\comcbbar$, couple together via the self-interference channel $\mH$.
Off-the-shelf codebooks for conventional half-duplex \mmwave systems (e.g., conjugate beams, \gls{dft} codebooks) do not necessarily offer much robustness to self-interference \cite{roberts_wcm}.
% Each transmit-receive beam pair will couple uniquely together, leading to various levels of self-interference.
% Those well-aligned with the self-interference channel are prone to inflict and couple self-interference while those poorly aligned will naturally reject self-interference.
% Since transmit and receive beams are coupled through the self-interference channel, the two 
% Intuitively, it may seem that conventional codebooks would have beams that resist self-interference given the highly directional beams that dense \mmwave arrays can produce.
% While somewhat accurate, this is not true for two key reasons: (i) transmit and receive beams are coupled via self-interference channel $\mH$, meaning the orthogonality of interest is actually that between the beams and the channel rather than the beams themselves; and (ii) the highly directional beams we are accustomed to are ``highly directional'' in the far-field but are not necessarily so in the near-field, where the transmit and receive arrays of a full-duplex device likely live.
Since even small amounts of self-interference can prohibit full-duplex operation, we are motivated us to design codebooks $\precbbar$ and $\comcbbar$ \textit{with full-duplex in mind}, such that self-interference is better rejected across all transmit-receive beam pairs while maintaining high beamforming gain.
% A full-duplex system making use of our designed codebooks would incur less self-interference while simultaneously delivering high beamforming gain and supporting beam alignment.

To justify the construction of these codebooks, we assume the self-interference channel $\mH$---little of which is currently known---can be estimated accurately and is sufficiently static over long periods.
We conduct our design under the assumption of perfect knowledge of $\mH$ and reserve future work for addressing imperfect channel knowledge.
Practically, we envision conducting thorough estimation of $\mH$ and execution of our design during initial setup of the transceiver and performing updates to the estimate of $\mH$ and to our design.

\subsection{Quantifying our Codebook Design Criteria}

% High beamforming gain afforded by dense analog beamforming arrays is essential for wide area \mmwave communication.
% Therefore, 
It is critical that our analog beamforming codebook offer high beamforming gain across a desired coverage region to combat severe \mmwave path loss .
% \begin{definition}[Transmit and receive coverage regions]
Let $\dirtxset$ be a set of $\Mtx$ transmit directions---or a \textit{transmit coverage region}---that our codebook aims to transmit toward, and let $\dirtxset$ be a set of $\Mrx$ receive directions---or a \textit{receive coverage region}---that our codebook aims to receive from.
\begin{gather}
\dirtxset = \braces{\dirtx\idx{i} : i = 1, \dots, \Mtx} \\
\dirrxset = \braces{\dirrx\idx{j} : j = 1, \dots, \Mrx}
\end{gather}
% \end{definition}
%\begin{figure}
%    \centering
%    \includegraphics[width=\linewidth,height=0.15\textheight,keepaspectratio]{plots/v01/generate_v01_process_v01_fig_00}
%    \caption{Example transmit and receive coverage regions $\dirtxset$ and $\dirrxset$ comprised of $\Mtx = \Mrx = 45$ points spaced uniformly in azimuth from $-60^\circ$ to $60^\circ$ with $15^\circ$ spacing and in elevation from $-30^\circ$ to $30^\circ$ with $15^\circ$ spacing, shown as their respective projections onto the $y$-$z$ plane, where looking into the page represents steering outward from each array.}
%    % \caption{Example discretized transmit and receive coverage regions $\dirtxset$ and $\dirrxset$ comprised of $\Mtx = \Mrx = 45$ points spaced uniformly in azimuth from $-60^\circ$ to $60^\circ$ with $15^\circ$ spacing and in elevation from $-30^\circ$ to $30^\circ$ with $15^\circ$ spacing. The directions $\dirtx\idx{i}$ and $\dirrx\idx{j}$ are shown here as their respective projections onto the $y$-$z$ plane, where looking into the page represents steering outward from each array (the positive $x$ direction).}
%    \label{fig:dirset}
%\end{figure}
% Corresponding to each of these transmit and receive directions, 
% Let $\atx{\dirtx} \in \setvectorcomplex{\Nt}$ be the transmit array response vector in the direction of $\aod$, and let $\arx{\aoa} \in \setvectorcomplex{\Nr}$ be the receive array response vector in the direction of $\aoa$.
Let $\Atx \in \setmatrixcomplex{\Nt}{\Mtx}$ be the matrix of transmit array response vectors evaluated at the directions in $\dirtxset$, and let $\Arx \in \setmatrixcomplex{\Nr}{\Mrx}$ be the matrix of receive array response vectors evaluated at the directions in $\dirrxset$.
\begin{align}
\Atx &= 
\begin{bmatrix}
\atx{\dirtx\idx{1}} & \atx{\dirtx\idx{2}} & \cdots & \atx{\dirtx\idx{\Mtx}}
\end{bmatrix}  \\
\Arx &= 
\begin{bmatrix}
\arx{\dirrx\idx{1}} & \arx{\dirrx\idx{2}} & \cdots & \arx{\dirrx\idx{\Mrx}}
\end{bmatrix}
\end{align}
%\begin{align}
%\Arx = \blkdiag{\arx{\aoa_1}, \arx{\aoa_2}, \dots, \arx{\aoa_{\Gr}}} \in \setmatrixcomplex{\Nr\Gr}{\Gr}
%\end{align}

We index the $\Mtx$ beams in $\precbbar$ and the $\Mrx$ beams in $\comcbbar$ according to $\dirtxset$ and $\dirrxset$, respectively, as $\precbbar = \braces{\vf_1, \dots, \vf_{\Mtx}}$ and $\comcbbar = \braces{\vw_1, \dots, \vw_{\Mrx}}$, 
%\begin{gather}
%\precbbar = \braces{\vf_1, \dots, \vf_{\Mtx}} \\
%\comcbbar = \braces{\vw_1, \dots, \vw_{\Mrx}}
%\end{gather}
where $\vf_i \in \precb$ is responsible for transmitting toward $\dirtx\idx{i}$ and $\vw_j \in \comcb$ is responsible for receiving from $\dirrx\idx{j}$.
Using this notation, we build the codebook matrix $\precbmat$ by stacking transmit beams $\vf_i$ as follows and the codebook matrix $\comcbmat$ analogously.
\begin{gather}
\precbmat = 
\begin{bmatrix}
\vf_1 & \vf_2 & \cdots & \vf_{\Mtx}
\end{bmatrix} \in \setmatrixcomplex{\Nt}{\Mtx} \\
\comcbmat = 
\begin{bmatrix}
\vw_1 & \vw_2 & \cdots & \vw_{\Mrx}
\end{bmatrix} \in \setmatrixcomplex{\Nr}{\Mrx}
\end{gather}

The coupling between the $i$-th transmit beam $\vf_i$ and the $j$-th receive beam $\vw_j$ can be captured by the product $\vw_j\ctrans \mH \vf_i$, which can be extended to all beam pairs as $\comcbmat\ctrans \mH \precbmat \in \setmatrixcomplex{\Mrx}{\Mtx}$.
Our design aims to minimize the \textit{average} coupling between beam pairs, which we denote $E$ and write as
% To reduce the coupling across beam pairs, our design aims to minimize the \textit{average} energy of $\comcbmat\ctrans \mH \precbmat$, which we denote $E$ and write as follows.
\begin{align}
E = \frac{1}{\Mtx\Mrx} \cdot \normfro{\comcbmat\ctrans \mH \precbmat}^2 \label{eq:power-si}.
\end{align}
% We are motivated to minimize the average coupling to ensure that low
In our design's attempt to minimize $E$, we must also ensure that the beams in $\precbmat$ and $\comcbmat$ remain useful in serving users within the coverage region; of course, unconstrained, $\precbmat$ and $\comcbmat$ minimize $E$ when driven to $\zeromat$.
Notice that our definition of $E$ neglects the large-scale channel gain $G$ and transmit gain $P$, which would merely scale $E$.
We now quantify a means to constrain our design to ensure it can maintain service.

Let $\Gtxmax^2 = \Nt^2$ and $\Grxmax^2 = \Nr^2$ be the maximum transmit and receive power gain possible by our arrays, which can be achieved toward $\dir$ with $\vf = \atx{\dir}$ and $\vw = \arx{\dir}$, respectively.
Let $\Gtx^2\parens{\dirtx\idx{i}}$ be the transmit power gain afforded by $\vf_i$ in the direction of $\dirtx\idx{i}$ and $\Grx^2\parens{\dirrx\idx{j}}$ be the receive power gain afforded by $\vw_j$ in the direction of $\dirrx\idx{j}$, which can be written as
\begin{align}
\Gtx^2\parens{\dirtx\idx{i}} &= \bars{\atx{\dirtx\idx{i}}\ctrans  \vf_i}^2 \leq \Gtxmax^2 \label{eq:Gtx} \\
\Grx^2\parens{\dirrx\idx{j}} &= \bars{\arx{\dirrx\idx{j}}\ctrans  \vw_j}^2 \leq \Grxmax^2 \label{eq:Grx}
\end{align}

%\begin{figure*}
%    \centering
%    \subfloat[A conventional codebook.]{\includegraphics[width=0.47\linewidth,height=0.25\textheight,keepaspectratio]{plots/v01/generate_v01_process_v01_fig_09_tgt-1}
%        \label{fig:matrix-cbf}}
%    \quad
%    \subfloat[Our codebook.]{\includegraphics[width=0.47\linewidth,height=0.25\textheight,keepaspectratio]{plots/v01/generate_v01_process_v01_fig_09_ours}
%        \label{fig:matrix-ours}}
%    \caption{The magnitude squared (in dB) of the beam coupling matrix $\comcbmat\ctrans \mH \precbmat$ for a conventional codebook and a codebook output by our design. Each transmit and receive codebook contains $45$ beams.}
%    \label{fig:matrix}
%\end{figure*}

%\begin{definition}[Minimum required transmit and receive beamforming gain]
%	Let $\Gtxmin^2$ and $\Grxmin^2$ be the minimum transmit and receive beamforming power gain for any precoder or combiner, respectively, in the direction it intends to serve; that is to say that $\prerfcbbar$ and $\comrfcbbar$ are required to have beams that satisfy
%	\begin{align}
%		\Gtxell \geq \Gtxmin \ \forall \ \ell \in \braces{1,\dots,\Mt} \label{eq:coverage-constraint-tx} \\
%		\Grxell \geq \Grxmin \ \forall \ \ell \in \braces{1,\dots,\Mr} \label{eq:coverage-constraint-rx}
%	\end{align}
%\end{definition}

To offer our design flexibility in reducing $E$, we tolerate some loss in beamforming gain toward our desired directions; in other words, we will aim for some beamforming gain less than $\Gtxmax^2$ and $\Grxmax^2$ in exchange for improved self-interference rejection.
% The beamforming gains we will aim for we term our \textit{target} transmit and receive beamforming gain.
% \begin{definition}[Target transmit and receive gain.]
    Let $\Gtxtgt^2$ and $\Grxtgt^2$ be the following \textit{target} transmit and receive beamforming gains for any transmit beam $\vf_i$ or receive beam $\vw_j$, respectively, in the direction it intends to serve. 
    % In other words, we desire
    \begin{gather}
    % \Gtx^2\parens{\dirtx\idx{i}} \approx 
    \Gtxtgt^2 = \Deltatx^2 \cdot \Gtxmax^2 \label{eq:target-gain-tx-approx} \\
    % \Grx^2\parens{\dirrx\idx{j}} \approx 
    \Grxtgt^2 = \Deltarx^2 \cdot \Grxmax^2 \label{eq:target-gain-rx-approx}
    % \Gtxtgt^2 \approx \Gtx^2\parens{\dirtx\idx{i}} \ \forall \ i \in \braces{1,\dots,\Mtx} \label{eq:target-gain-tx-approx} \\
    % \Grxtgt^2 \approx \Grx^2\parens{\dirrx\idx{j}} \ \forall \ j \in \braces{1,\dots,\Mrx} \label{eq:target-gain-rx-approx}
    \end{gather}
    Here, $0 < \Deltatx^2 \leq 1$ and $0 < \Deltarx^2 \leq 1$ are tolerated losses in beamforming gain (design parameters) that can be chosen by system engineers.
    % which we will more precisely quantify shortly.
    Note that specific target gains for each beam in each codebook can be defined uniquely, though we use a common target codebook-wide for simplicity.
    % Let $\Gtxtgt^2$ and $\Grxtgt^2$ be defined conveniently using $0 < \Deltatx^2 \leq 1$ and $0 < \Deltarx^2 \leq 1$ as follows.
    % \begin{align}
    % \Gtxtgt^2 = \Deltatx^2 \cdot \Gtxmax^2 \\
    % \Grxtgt^2 = \Deltarx^2 \cdot \Grxmax^2
    % \end{align}
    % $\Deltatx^2$ and $\Deltarx^2$ capture the loss in beamforming gain that we tolerate as part of our design.
    %	More convenient log-scale expressions are
    %	\begin{align}
    %	\todB{\Gtxtgt^2} = \todB{\Gtxmax^2} + \todB{\Deltatx^2} \\
    %	\todB{\Grxtgt^2} = \todB{\Grxmax^2} + \todB{\Deltarx^2}
    %	\end{align}
    %	where $\todB{\Deltatx^2} \leq 0$ and $\todB{\Deltarx^2} \leq 0$.
    % The tolerated losses $\Deltatx^2$ and $\Deltarx^2$ are design parameters that can be determined by system engineers.
% \end{definition}

% \begin{definition}[Transmit and receive gain variance.]
    To quantify each transmit and receive beam approximately meeting the target gains in \eqref{eq:target-gain-tx-approx} and \eqref{eq:target-gain-rx-approx}, we introduce a variance tolerance for each.
    Let $\sigmatx^2 \geq 0$ be the maximum variance tolerated in achieving the target transmit (magnitude) gain  $\Gtxtgt$ across beams in the transmit codebook $\precbbar$, normalized to $\Gtxtgt^2$.
    Defining $\sigmarx^2 \geq 0$ analogously, we have
        \begin{gather}
    \frac{1}{\Mtx} \cdot \sum_{i=1}^{\Mtx} \frac{\bars{\Gtxtgt - \Gtx\parens{\dirtx\idx{i}}}^2}{\Gtxtgt^2} \leq \sigmatx^2 \label{eq:sigma-tx} \\
    \frac{1}{\Mrx} \cdot \sum_{j=1}^{\Mrx} \frac{\bars{\Grxtgt - \Grx\parens{\dirrx\idx{j}}}^2}{\Grxtgt^2} \leq \sigmarx^2  \label{eq:sigma-rx}.
    \end{gather}
%    \begin{align}
%    \frac{1}{\Mtx} \cdot \sum_{i=1}^{\Mtx} \frac{\bars{\Gtxtgt - \Gtx\parens{\dirtx\idx{i}}}^2}{\Gtxtgt^2} \leq \sigmatx^2 \label{eq:sigma-tx}
%    \end{align}
%    Similarly, let $\sigmarx^2 \geq 0$ be the maximum variance tolerated in achieving the target receive (magnitude) gain $\Grxtgt$ across all beams in the receive codebook $\comcbbar$, normalized to $\Grxtgt^2$.
%    \begin{align}
%    \frac{1}{\Mrx} \cdot \sum_{j=1}^{\Mrx} \frac{\bars{\Grxtgt - \Grx\parens{\dirrx\idx{j}}}^2}{\Grxtgt^2} \leq \sigmarx^2  \label{eq:sigma-rx}
%    \end{align}
    % Design parameters $\Deltatx^2$, $\Deltarx^2$, $\sigmatx^2$ and $\sigmarx^2$ give engineers the ability to tune our methodology based on system requirements.
    % More variance across transmit and receive gain provided by our beams may increase the disparity of service seen across users but would also allow our design more flexibility in its attempt to reduce self-interference.
% \end{definition}

\subsection{Assembling our Codebook Design Problem}
With expressions for our codebook design criteria in hand, we now turn our attention to assembling a formal design problem.
Let $\gtxtgt = \Gtxtgt \cdot \onevec{\Mtx}$ and $\grxtgt = \Grxtgt \cdot \onevec{\Mrx}$ be vectors containing the target transmit and receive gains at each entry.
%\begin{align}
%\gtxtgt = \Gtxtgt \cdot \onevec{\Mtx} \label{eq:gtx} \\
%\grxtgt = \Grxtgt \cdot \onevec{\Mrx} \label{eq:grx}
%\end{align}
Using this, the expressions of \eqref{eq:sigma-tx} and \eqref{eq:sigma-rx} can be written equivalently (up to an arbitrary phase shift of $\vf_i$ or $\vw_j$) as follows, which we refer to as our \textit{coverage constraints},
\begin{align}
\normtwo{\gtxtgt - \diag{\Atx\ctrans \precbmat}}^2 \leq \sigmatx^2 \cdot \Gtxtgt^2 \cdot \Mtx \label{eq:coverage-constraint-tx} \\
\normtwo{\grxtgt - \diag{\Arx\ctrans \comcbmat}}^2 \leq \sigmarx^2 \cdot \Grxtgt^2 \cdot \Mrx \label{eq:coverage-constraint-rx}
\end{align}
where the $(i,i)$-th entry of $\Atx\ctrans \precbmat$ has magnitude $\Gtx\parens{\dirtx\idx{i}}$ and the $(j,j)$-th entry of $\Arx\ctrans \comcbmat$ has magnitude $\Grx\parens{\dirrx\idx{j}}$.
By satisfying \eqref{eq:coverage-constraint-tx} and \eqref{eq:coverage-constraint-rx}, we can ensure that our codebooks $\precbbar$ and $\comcbbar$ adequately serve our coverage regions.
% Note that these two constraints capture the gain and coverage constraints of our two codebooks $\prerfcbbar$ and $\comrfcbbar$.

% The parameters $\Gtxtgt$ and $\Grxtgt$ (equivalently $\Deltatx$ and $\Deltatx$) along with $\sigmatx^2$ and $\sigmatx^2$ can be tuned as desired by system engineers. 
% We will explore their impact on the design our numerical results in \secref{sec:simulation-results}.
%Convenient log-scale representations of $\sigmatx^2$ and $\sigmarx^2$ are 
%\begin{align}
%\todB{\sigmatx^2} = 10 \cdot \logten{\sigmatx^2} \\
%\todB{\sigmatx^2} = 10 \cdot \logten{\sigmatx^2} 
%\end{align}
%where it practical to take $\todB{\sigmatx^2} \leq 0$ dB (i.e., $\sigmatx^2 \leq 1$) and $\todB{\sigmatx^2} \leq 0$ dB (i.e., $\sigmatx^2 \leq 1$).
%Lower $\sigmatx^2$, $\sigmatx^2$ results in stricter constraints (less variance in achieving targets).

Using these coverage constraints, we formulate the following design problem to reduce $E$ while maintaining coverage and satisfying quantized phase and amplitude control.
\begin{subequations} \label{eq:problem-full-1}
    \begin{align}
    \min_{\precbmat,\comcbmat} &\ \normfro{\comcbmat\ctrans \mH \precbmat}^2 \label{eq:problem-objective-1} \\
    \st 
    &\ \normtwo{\gtxtgt - \diag{\Atx\ctrans \precbmat}}^2 \leq \sigmatx^2 \cdot \Gtxtgt^2 \cdot \Mtx \label{eq:problem-coverage-tx-1} \\
    &\ \normtwo{\grxtgt - \diag{\Arx\ctrans \comcbmat}}^2 \leq \sigmarx^2 \cdot \Grxtgt^2 \cdot \Mrx \label{eq:problem-coverage-rx-1} \\
    &\ \entry{\precbmat}{:,i} \in \precb \ \forall \ i = 1, \dots, \Mtx \label{eq:problem-quantize-tx-1} \\
    &\ \entry{\comcbmat}{:,j} \in \comcb \ \forall \ j = 1, \dots, \Mrx \label{eq:problem-quantize-rx-1} 
    \end{align}
\end{subequations}
% The parameters $\Gtxtgt$ and $\Grxtgt$ (equivalently $\Deltatx$ and $\Deltarx$) along with $\sigmatx^2$ and $\sigmarx^2$ can be tuned as desired by system engineers.
Decreasing $\Gtxtgt^2$ and $\Grxtgt^2$ will relax the constraints of our design, allowing it to better reduce self-interference coupled by our transmit and receive beams; this helps facilitate full-duplexing service to two devices but degrades the service each device experiences.
Increasing $\sigmatx^2$ and $\sigmarx^2$ will increase the flexibility of our design but weakens the coverage guarantee.
% We examine this tradeoff in our numerical results in \secref{sec:simulation-results}.

\comment{
\subsection{Capturing Self-Interference Channel Errors}
Abiding by the scope of this work, it is reasonable to assume the self-interference channel $\mH$ may not be perfectly estimated and may vary slowly over time.
Small changes in $\mH$ may render designs of $\precbmat$ and $\comcbmat$ far less useful if not designed for such, especially since self-interference is likely overwhelmingly strong.
Suppose instead of having perfect knowledge of true over-the-air channel, $\mH$ is an estimate of the true channel $\bar{\mH}$ and knowledge of an estimation error bound.
Specifically, let us define the set of all possible $\bar{\mH}$ as
\begin{align}
\Hsiset = \braces{\bar{\mH} = \mH + \sqrt{\Nt \Nr} \cdot \mDelta : \normfro{\mDelta} \leq \epsilon}
\end{align}
where the matrix $\mDelta$ captures the error in our estimate $\mH$ and is bounded by its Frobenius norm as $\normfro{\mDelta} \leq \epsilon$.
Note that other norm-bound errors can be encapsulated by appropriate choice of $\epsilon$ via norm equivalence.
We take a worst-case optimization approach to design $\precbmat$ and $\comcbmat$ in the presence of channel error $\mDelta$, leading to problem \eqref{eq:problem-full-2}.
\begin{subequations} \label{eq:problem-full-2}
    \begin{align}
    \min_{\precbmat,\comcbmat} &\ \max_{\normfro{\mDelta} \leq \epsilon} \ \normfro{\comcbmat\ctrans \parens{\mH + \sqrt{\Nt \Nr} \cdot \mDelta} \precbmat} \label{eq:problem-objective-2} \\
    \st 
    &\ \normtwo{\gtxtgt - \diag{\Atx\ctrans \precbmat}}^2 \leq \sigmatx^2 \cdot \Gtxtgt^2 \cdot \Mtx \label{eq:problem-coverage-tx-2} \\
    &\ \normtwo{\grxtgt - \diag{\Arx\ctrans \comcbmat}}^2 \leq \sigmarx^2 \cdot \Grxtgt^2 \cdot \Mrx \label{eq:problem-coverage-rx-2} \\
    &\ \entry{\precbmat}{:,i} \in \precb \ \forall \ i = 1, \dots, \Mtx \label{eq:problem-quantize-tx-2} \\
    &\ \entry{\comcbmat}{:,j} \in \comcb \ \forall \ j = 1, \dots, \Mrx \label{eq:problem-quantize-rx-2} 
    \end{align}
\end{subequations}
When $\normfro{\mDelta} \leq \epsilon$, the objective
\begin{align}
\normfro{\comcbmat\ctrans \parens{\mH + \sqrt{\Nt \Nr} \cdot \mDelta} \precbmat}
\end{align} is upper bounded by
% The objective in \eqref{eq:problem-objective-2} is upper bounded by 
\begin{align}
% \max_{\normfro{\mDelta} \leq \epsilon} \normfro{\comcbmat\ctrans \parens{\mH + \sqrt{\Nt \Nr} \cdot \mDelta} \precbmat}
% &= \normfro{\comcbmat\ctrans \mH \precbmat} + \normfro{\comcbmat\ctrans \mDelta \precbmat} \\
% &\leq 
\normfro{\comcbmat\ctrans \mH \precbmat} + \epsilon \cdot \sqrt{\Nt \Nr} \cdot \normtwo{\comcbmat} \cdot \normtwo{\precbmat}
\end{align}
which does not depend on $\mDelta$ but merely $\epsilon$.
By this fact, problem \eqref{eq:problem-full-2} can be reformulated to the slightly different problem \eqref{eq:problem-full-3} as follows, which takes a familiar regularizing form.
Instead of using $\epsilon$ directly, we use $\tilde{\epsilon} \leq \epsilon$, which can be chosen to better handle general channel errors, rather than strictly the worst-case one.
\begin{subequations} \label{eq:problem-full-3}
    \begin{align}
    \min_{\precbmat,\comcbmat} &\ \normfro{\comcbmat\ctrans \mH \precbmat} + \tilde{\epsilon} \cdot \sqrt{\Nt \Nr} \cdot \normtwo{\precbmat} \cdot \normtwo{\comcbmat} \label{eq:problem-objective-3} \\
    \st 
    &\ \normtwo{\gtxtgt - \diag{\Atx\ctrans \precbmat}}^2 \leq \sigmatx^2 \cdot \Gtxtgt^2 \cdot \Mtx \label{eq:problem-coverage-tx-3} \\
    &\ \normtwo{\grxtgt - \diag{\Arx\ctrans \comcbmat}}^2 \leq \sigmarx^2 \cdot \Grxtgt^2 \cdot \Mrx \label{eq:problem-coverage-rx-3} \\
    &\ \entry{\precbmat}{:,i} \in \precb \ \forall \ i = 1, \dots, \Mtx \label{eq:problem-quantize-tx-3} \\
    &\ \entry{\comcbmat}{:,j} \in \comcb \ \forall \ j = 1, \dots, \Mrx \label{eq:problem-quantize-rx-3} 
    \end{align}
\end{subequations}
Problem \eqref{eq:problem-full-3} is much more straightforwardly solved than problem \eqref{eq:problem-full-2} since it does not require optimization over $\mDelta$.
By solving problem \eqref{eq:problem-full-3}, codebooks $\precbbar$ and $\comcbbar$ will contain beams that are more robust to channel errors due to estimation or slight variations over time.
Note that problem \eqref{eq:problem-full-1} and problem \eqref{eq:problem-full-3} are identical when $\tilde{\epsilon} = 0$.
}

% \input{alg/algorithm-gradient-step.tex}

% \input{alg/algorithm-full-i.tex}

% \input{sec-contribution-i.tex}

% \input{sec-contribution-i-old.tex}

% \pagebreak

\subsection{Solving our Codebook Design Problem} % \label{sec:contribution-ii}

Having arrived at our design problem \eqref{eq:problem-full-1}, we now set out to solve for our design.
The non-convexity posed by digitally-controlled phase shifters and attenuators, captured by \eqref{eq:problem-quantize-tx-1} and \eqref{eq:problem-quantize-rx-1}, presents difficulty in solving this problem.
%  problem \eqref{eq:problem-full-3} is non-convex due to the limited resolution of our digitally-controlled phase shifters and attenuators used for analog beamforming, captured by \eqref{eq:problem-quantize-tx-3} and \eqref{eq:problem-quantize-rx-3}.
% \edit{Is the objective jointly convex?}
In general, we found that we cannot handle this non-convexity by merely ignoring it, solving, and then projecting the solutions $\precbmat\opt$ and $\comcbmat\opt$ onto $\precb$ and $\comcb$, respectively, though this may offer reasonable results when $\bitsphase$ and $\bitsamp$ are sufficiently high.
To present a general approach to handling this non-convexity, we present the following alternating minimization with the understanding that more sophisticated algorithms can only improve the results we observe.

We define $\project{\mA}{\mathcal{A}}$ as the projection of the elements of $\mA$ onto the set $\mathcal{A}$.
Our algorithm begins by initializing $\precbmat$ and $\comcbmat$ as the projections of scaled $\Atx$ and $\Arx$ onto $\precb$ and $\comcb$, respectively, quantizing their entries to have phase and amplitude resolutions $\bitsphase$ and $\bitsamp$.
That is, $\precbmat \leftarrow \project{\Atx \cdot \Deltatx}{\precb}$ and $\comcbmat \leftarrow \project{\Arx \cdot \Deltarx}{\comcb}$, which initializes our beams to achieve gains $\Gtxtgt^2$ and $\Grxtgt^2$ across $\dirtxset$ and $\dirrxset$, respectively.
%\begin{gather}
%\precbmat \leftarrow \project{\Atx \cdot \Deltatx}{\precb} \\
%\comcbmat \leftarrow \project{\Arx \cdot \Deltarx}{\comcb} 
%\end{gather}
% This 

In this approach, we approximately solve problem \eqref{eq:problem-full-1} by solving for the codebooks beam-by-beam. 
We separate problem \eqref{eq:problem-full-1} into problems \eqref{eq:problem-full-4a} and \eqref{eq:problem-full-4b} below, which solve for the $i$-th transmit beam and $j$-th receive beam, respectively.
Problems \eqref{eq:problem-full-4a} and \eqref{eq:problem-full-4b} are convex since they remove the hardware constraints and can be readily solved using a convex solver (we used \cite{cvx}).
We solve for the $i$-th transmit beam $\vf_i$ in problem \eqref{eq:problem-full-4a}, starting with $i=1$ and the initialized $\comcbmat$.
% Note that we have introduced the constraint \eqref{eq:problem-entry-tx-4a} to ensure our beamforming weights can be implemented using attenuators.
\begin{subequations} \label{eq:problem-full-4a}
	\begin{align}
	\min_{\vf_i} &\ \normtwo{\comcbmat\ctrans \mH \vf_i} \label{eq:problem-objective-4a} \\
	\st 
	&\ \bars{\Gtxtgt - \atx{\dirtx\idx{i}}\ctrans \vf_i}^2 \leq \sigmatx^2 \cdot \Gtxtgt^2 \label{eq:problem-coverage-tx-4a} \\
	&\ \bars{\entry{\vf_i}{n}} \leq 1 \ \forall \ n = 1, \dots, \Nt \label{eq:problem-entry-tx-4a}
	\end{align}
\end{subequations}
Then, the solution $\vf_i\opt$ is projected onto the set $\precb$ before updating it in the matrix $\precbmat$.
\begin{align}
\entry{\precbmat}{:,i} \leftarrow \project{\vf_i\opt}{\precb}
\end{align}
The updated $\precbmat$ is then used when solving for the $j$-th receive beam (beginning with $j=1$) in problem \eqref{eq:problem-full-4b}.
\begin{subequations} \label{eq:problem-full-4b}
	\begin{align}
	\min_{\vw_j} &\ \normtwo{\vw_j\ctrans \mH \precbmat}  \label{eq:problem-objective-4b} \\
	\st 
	&\ \bars{\Grxtgt - \arx{\dirrx\idx{j}}\ctrans \vw_j}^2 \leq \sigmarx^2 \cdot \Grxtgt^2 \label{eq:problem-coverage-rx-4b} \\
	&\ \bars{\entry{\vw_j}{n}} \leq 1 \ \forall \  = 1, \dots, \Nr \label{eq:problem-entry-rx-4b} 
	\end{align}
\end{subequations}
The solution $\vw_j\opt$ is then projected onto $\comcb$ before updating $\comcbmat$.
\begin{align}
\entry{\comcbmat}{:,j} \leftarrow \project{\vw_j\opt}{\comcb}
\end{align}
We iteratively solve for the $(i+1)$-th transmit beam and then the $(j+1)$-th receive beam until all $\Mtx$ and $\Mrx$ beams are solved for.
% When $\Mtx = \Mrx$, the transmit and receive beams can be solved one after another.
When $\Mtx \neq \Mrx$, the remaining transmit/receive beams can be solved for after $\minop{\Mtx,\Mrx}$ iterations.
We have found that this iterative approach handles the quantization of $\precbmat$ and $\comcbmat$ quite well by \textit{progressively} incorporating quantization beam-by-beam, allowing later beams to account for the quantization imposed on earlier beams.

\section{Numerical Results} \label{sec:simulation-results}

% -------------------------------------------------------------------------------------------------
% Setup.
% -------------------------------------------------------------------------------------------------
To evaluate our design, we simulated a simple $30$ GHz network where a full-duplex transceiver transmits to a downlink user and receives from an uplink user.
The two users are equipped with a single antenna while the full-duplex device has two $8 \times 8$ half-wavelength \upas one for transmission and one for reception.
We drew realizations of our network in a Monte Carlo fashion.
The channel vectors $\vhtx \in \setmatrixcomplex{64}{1}$ to the downlink user and $\vhrx \in \setvectorcomplex{64}$ to the uplink user are simulated as \gls{los} channels with normally distributed gain, where users are distributed uniformly across the transmit/receive coverage regions at each realization.
% The choice of a particular self-interference channel model is a difficult one to justify practically given there lacks a well-accepted, measurement-backed model.
For the self-interference channel $\mH$, we consider the spherical-wave channel model \cite{spherical_2005}, which captures idealized near-field interaction between the transmit and receive arrays of the full-duplex device, described as
\begin{align}
\entry{\mH}{m,n} = \frac{\gamma}{r_{n,m}}\exp \left(-\j 2 \pi \frac{r_{n,m}}{\lambda} \right)
\end{align}
where $r_{n,m}$ is the distance between the $n$-th transmit antenna and the $m$-th receive antenna, $\lambda$ is the carrier wavelength, and $\gamma$ is a normalizing factor to satisfy $\normfro{\mH}^2 = \Nt\Nr$.
To realize such a channel, we have separated our transmit and receive arrays by $10 \lambda$ in the azimuth plane.
While this model may not hold in practice, it provides us a sensible starting point to evaluate our design when the self-interference channel is dominated by near-field interaction.
% In addition, it motivates the possibility that our design be executed on some deterministic or well estimated near-field model while treating imperfections (e.g., stemming from far-field reflections) as channel estimation error.

% As such, we consider two channel models: (i) a Rayleigh-faded channel and (2) an idealized near-field model.
% To ensure the channels are compared fairly, they each satisfy $\ev{\normfro{\mH}^2} = \Nt \Nr$, meaning only their spatial characteristics dictate our design's ability to reduce the coupling of self-interference, rather than the energy contained therein.

Our transmit and receive coverage regions are comprised of uniformly spaced points in azimuth from $-60^\circ$ to $60^\circ$ with $15^\circ$ spacing and in elevation from $-30^\circ$ to $30^\circ$ with $15^\circ$ spacing.
This amounts to $\Mtx = \Mrx = 45$ total directions in $\dirtxset$ and $\dirrxset$.
We have assumed the analog beamforming networks use log-stepped attenuators with $0.25$ dB of attenuation per \gls{lsb}.

% -------------------------------------------------------------------------------------------------
% Metrics.
% -------------------------------------------------------------------------------------------------
% \subsection{Metrics}
We assess our design with sum spectral efficiency of the transmit and receive links.
Suppose the transmit and receive links have \glspl{snr} $\snrtx$ and $\snrrx$, respectively, which capture the received powers relative to noise \textit{without} beamforming.
Transmitting with beamformer $\vf$, the downlink user can achieve a spectral efficiency
\begin{align}
	\setx &= \logtwo{1 + \frac{\snrtx}{\Nt} \cdot \bars{\vhtx\ctrans \vf}^2}
\end{align}
where $\Nt\inv$ handles power splitting in the transmit beamforming network.
Plagued by self-interference, the uplink user sees % the spectral efficiency
\begin{align}
	\serx &= \logtwo{1 + \frac{\snrrx \cdot \bars{\vw\ctrans \vhrx}^2}{\normtwo{\vw}^2 + \inr \cdot \bars{\vw\ctrans \mH \vf}^2}}
\end{align}
where $\bars{\vw\ctrans \mH \vf}^2$ is the self-interference coupling factor between the transmit and receive beams and $\inr = {P \cdot G^2}/{N_0}$
%\begin{align}
%	\inr = \frac{P \cdot G^2}{N_0}
%\end{align}
is the \gls{inr} of self-interference without beamforming; $N_0$ is the additive noise variance.
$\inr$ is an important quantity for evaluating this work since it captures the \textit{isolation} $G^{-2}$ between the transmit and receive arrays, allowing us to abstract it from the \textit{spatial} coupling of transmit and receive beams.
The $\inr$ of a \mmwave full-duplex system depends on a variety of factors including transmit power, noise power, and system setup.
The uplink and downlink spectral efficiencies are bounded by their Shannon capacities (neglecting self-interference on the uplink) with unconstrained beams $\vf$ and $\vw$.
The sum spectral efficiency $\setx + \serx$ is bounded by the full-duplex capacity $\capfd$, the sum of the two Shannon capacities.
Under equal \gls{tdd}, the half-duplex capacity is simply $\caphd = 0.5 \cdot \capfd$.
Note that the use of beamforming codebooks will naturally fall short of these capacities.

%\begin{align}
%	\sinrrx = \frac{\snrrx}{1 + \inr}
%\end{align}

%\begin{align}
%R = \setx + \serx
%\end{align}

% -------------------------------------------------------------------------------------------------
% Benchmarks.
% -------------------------------------------------------------------------------------------------
\begin{figure}
    \centering
    \includegraphics[width=\linewidth,height=0.22\textheight,keepaspectratio]{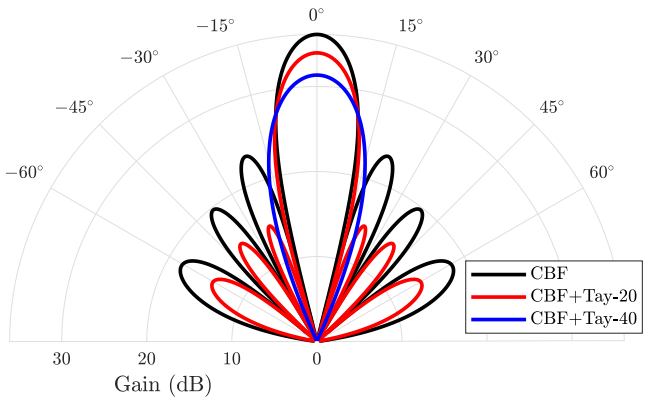}
    \caption{The azimuth cut of a broadside beam from each benchmark codebook.}
    \label{fig:benchmarks}
\end{figure}

Considering ours is the first known codebook design \textit{for \mmwave full-duplex}, we evaluate our codebook design against the following three conventional codebooks:
(i) conjugate beamforming (CBF) where $\vf_i = \atx{\dirtx\idx{i}}$ and $\vw_j = \arx{\dirrx\idx{j}}$; 
    % \begin{align}
    % \vf_i = \atx{\dirtx\idx{i}}, \ \vw_j = \arx{\dirrx\idx{j}}
    % \end{align}
%     \item % CBF plus Taylor windowing with $20$ dB of side lobe attenuation 
    (ii) CBF+Tay-20 where
    % \begin{align}
    % \vf_i = \atx{\dirtx\idx{i}} \odot \vv, \ \vw_j = \arx{\dirrx\idx{j}} \odot \vv \label{eq:cbf-tay}
    % \end{align}
    $\vf_i = \atx{\dirtx\idx{i}} \odot \vv$ and $\vw_j = \arx{\dirrx\idx{j}} \odot \vv$ 
    and $\vv$ is a Taylor window with $20$ dB of side lobe suppression applied element-wise; 
    and (iii) CBF+Tay-40, defined as in CBF+Tay-20, except $\vv$ is a Taylor window with $40$ dB of side lobe suppression.
    % \item CBF plus Taylor windowing with $40$ dB of side lobe attenuation (CBF+Tay-40) where $\vf_i = \atx{\dirtx\idx{i}} \odot \vv$ and $\vw_j = \arx{\dirrx\idx{j}} \odot \vv$ and $\vv$ is a Taylor window with $40$ dB of side lobe suppression.
% \end{enumerate}
% (i) (ii) ; and (iii) conjugate beamforming with Taylor windowing for $40$ dB of side lobe attenuation (CBF+Tay-40).
The beam pattern of each of our three codebooks is shown in \figref{fig:benchmarks}.
CBF offers maximum beamforming gain with a narrow main lobe but exhibits high side lobe levels.
Taylor windowing reduces the side lobe levels at the cost of lessened beamforming gain and a wider main lobe (e.g., CBF+Tay-20 and CBF+Tay-40 lose about $2$ dB and $5$ dB in beamforming gain, respectively).
% To compare these benchmarks to our design, we can trivially scale them according to $\Deltatx^2$ and $\Deltarx^2$, which fairly accounts for the beamforming loss we permit our codebook to tolerate.
% This will be more clear shortly.

\begin{figure}
	\centering
	\includegraphics[width=\linewidth,height=0.22\textheight,keepaspectratio]{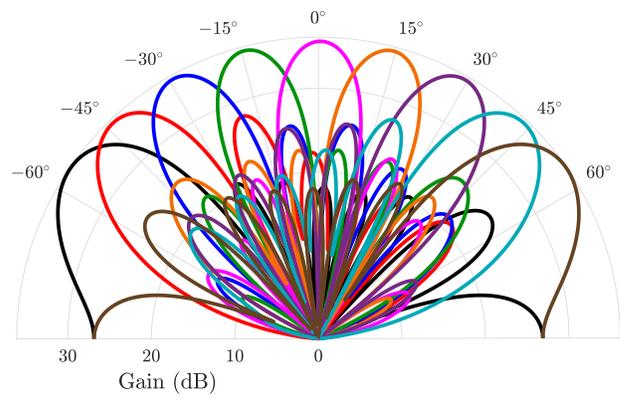}
	% \caption{The patterns of beams from our transmit codebook serving various azimuth directions at an elevation of $0^\circ$, where $\Deltatx^2 = \Deltarx^2 = 0$ dB, $\sigmatx^2 = \sigmarx^2 = -20$ dB, and $\bitsphase = \bitsamp = 5$ bits. The receive codebook looks nearly identical.}
    % \caption{The azimuth cuts of beams from our transmit codebook serving various azimuth directions at an elevation of $0^\circ$, where $\Deltatx^2 = \Deltarx^2 = 0$ dB, $\sigmatx^2 = \sigmarx^2 = -20$ dB, and $\bitsphase = \bitsamp = 5$ bits.}
     \caption{The azimuth cuts of beams from our transmit codebook serving various azimuth directions at an elevation of $0^\circ$.}
	\label{fig:patterns}
\end{figure}

% -------------------------------------------------------------------------------------------------
% Our beam patterns.
% -------------------------------------------------------------------------------------------------
Let us begin by visually inspecting the beams produced by our codebook design, which we illustrate in \figref{fig:patterns}.
We designed our codebooks using $\Deltatx^2 = \Deltarx^2 = 0$ dB and $\sigmatx^2 = \sigmarx^2 = -20$ dB, where we expect our beams to have nearly full transmit/receive gain and some flexibility in meeting this gain.
We use phase and attenuator resolutions $\bitsphase = \bitsamp = 5$ bits, which can be found in commercial phased arrays.
First, we note that our beams cover $-60^\circ$ to $60^\circ$ in $15^\circ$ steps as was specified by our transmit and receive coverage regions.
A user falling anywhere in the transmit/receive coverage region can confidently expect to see high gain.
Notice that our design appears to creatively shape side lobes, rather than merely shrinking them; this is its attempt to strategically cancel self-interference \textit{spatially}.
% Understanding the inter-user interference these side lobes may inflict in larger network settings will be important future work.

%---

\begin{figure}
	\centering
	\includegraphics[width=\linewidth,height=\textheight,keepaspectratio]{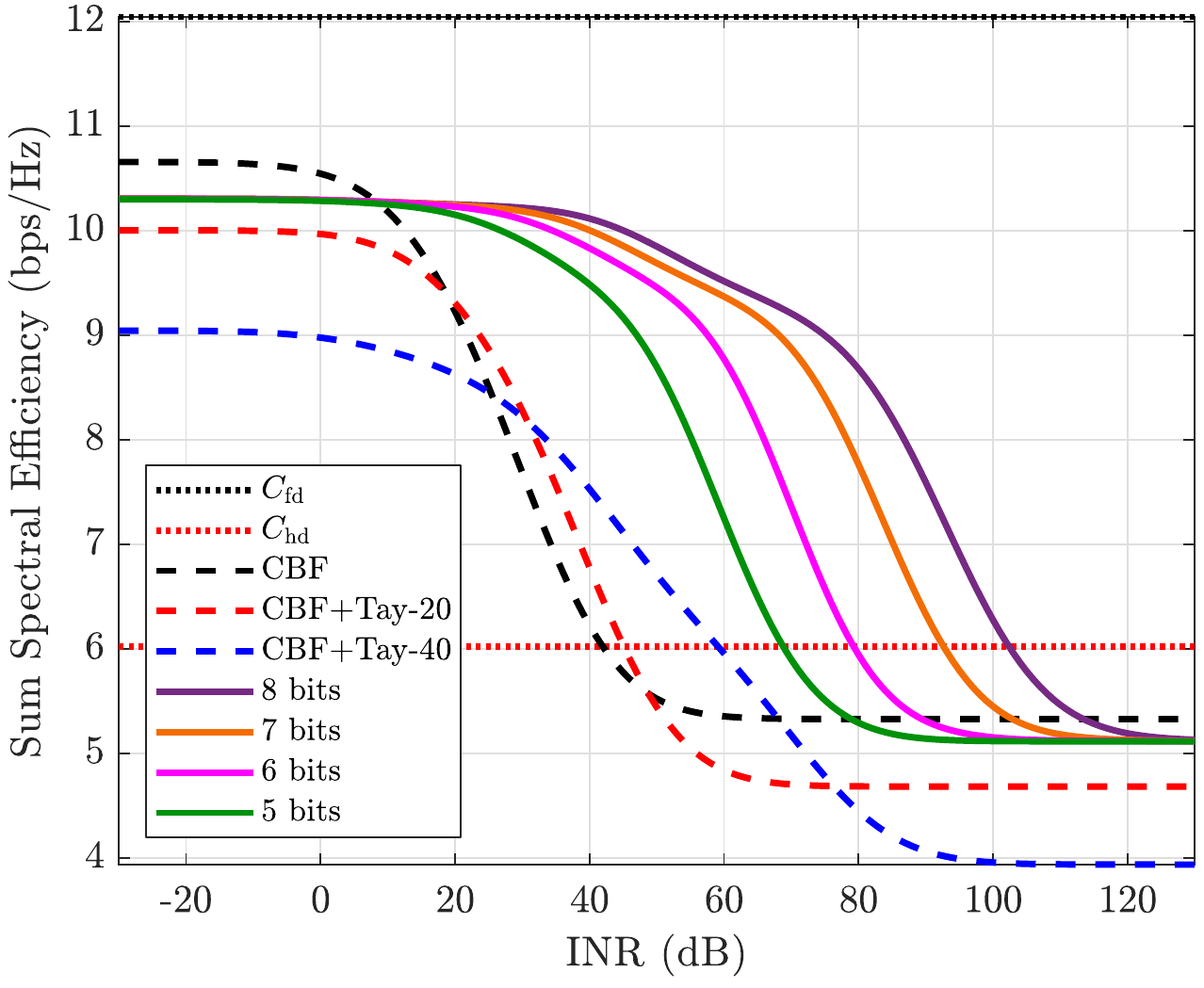}
	% \caption{Sum spectral efficiency $\setx + \serx$ as a function of $\inr$ for various codebooks, where $\Deltatx^2 = \Deltarx^2 = 0$ dB, $\sigmatx^2 = \sigmarx^2 = -20$ dB, and $\snrtx = \snrrx = 0$ dB.}
    \caption{Sum spectral efficiency $\setx + \serx$ as a function of $\inr$ for various codebooks, where $\snrtx = \snrrx = 0$ dB.}
	\label{fig:batman}
\end{figure}

Now, in \figref{fig:batman}, we evaluate the sum spectral efficiency $\setx + \serx$ offered by the considered benchmarks versus our codebook for various resolutions $\bitsphase = \bitsamp \in \braces{5,6,7,8}$ bits.
We begin by fixing $\snrtx = \snrrx = 0$ dB and varying $\inr$ from $-30$ dB to $130$ dB.
At $\inr \ll 0$ dB, the inherent isolation between the transmit and receive arrays is high, meaning the system can operate in full-duplex fashion, even with high coupling between transmit and receive beams.
This is clearly seen for all codebooks shown in \figref{fig:batman}, where they all maintain appreciable fractions of the full-duplex capacity at low $\inr$.
CBF beams offer maximal beamforming gain and thus the highest sum spectral efficiency at low $\inr$, with our design just below.
The other two benchmarks fall below due to their tradeoff of beamforming gain for side lobe suppression.
% Our design achieves nearly the beamforming gain of CBF.

As $\inr$ increases, the coupling between transmit and receive beams plays a significant role in what level of self-interference is present, and thus, what uplink spectral efficiency $\serx$ is achieved.
% Recall that downlink performance $\setx$ depends only on transmit beamforming gain, not self-interference (nor $\inr$).
Notice that the CBF codebook begins to taper off first as $\inr$ is increased, followed by CBF+Tay-20 and CBF+Tay-40 courtesy of their side lobe suppression.
% At around $\inr = 20$ dB, CBF+Tay-20 becomes the optimal codebook to use, and at around $\inr = 30$ dB, CBF+Tay-40 overtakes CBF+Tay-20.
% This is thanks to the Taylor windowing side lobe suppression, improving the robustness to self-interference.
Our design, on the other hand, is comparable or outperforms all three benchmarks across all $\inr$ thanks to its robustness to self-interference.
From $\inr = 10$ dB to $\inr = 80$ dB or more, our design outperforms all three conventional codebooks.
At a sum spectral efficiency of $8$ bps/Hz, we can see that our design offers $20$ dB of robustness to $\inr$ with $\bitsphase = \bitsamp = 5$ bits versus the conventional codebooks; with each added bit, we gain approximately $10$ dB of robustness.
% by offering high beamforming gain while \textit{simultaneously} reducing self-interference.
When $\inr \to 130$ dB, the system becomes overwhelmed with self-interference driving $\serx \to 0$ bps/Hz, even with our design's low coupling between transmit and receive beams.
% Clearly, more phase and attenuator resolution yields even better performance by relaxing design constraints.
% The improvement is quite significant from $5$ bits to $8$ bits, each bit adding around $10$ dB of $\inr$ robustness.

%\begin{figure}
%	\centering
%	\includegraphics[width=\linewidth,height=\textheight,keepaspectratio]{plots/v21/generate_v21_process_v02_fig_04.pdf}
%	\caption{Caption.}
%	\label{fig:}
%\end{figure}

\begin{figure}[!t]
	\centering
	\includegraphics[width=\linewidth,height=\textheight,keepaspectratio]{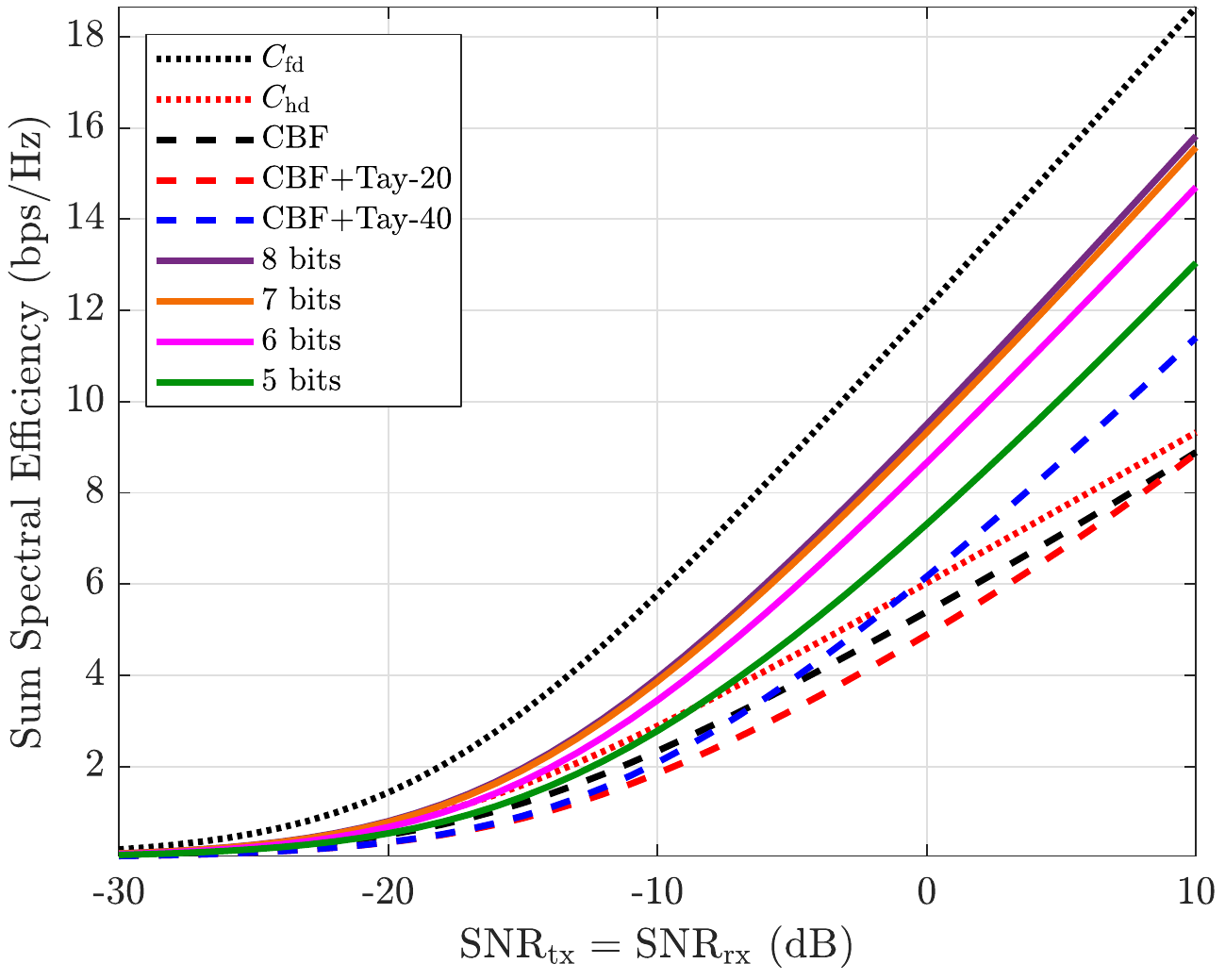}
	% \caption{Sum spectral efficiency $\setx + \serx$ as a function of $\snrtx = \snrrx$ for various codebooks, where $\Deltatx^2 = \Deltarx^2 = 0$ dB, $\sigmatx^2 = \sigmarx^2 = -20$ dB, and $\inr = 60$ dB.}
   	\caption{Sum spectral efficiency $\setx + \serx$ as a function of $\snrtx = \snrrx$ for various codebooks, where $\inr = 60$ dB.}
	\label{fig:ironman}
\end{figure}

Now, we consider \figref{fig:ironman}, where we fix $\inr = 60$ dB, vary $\snrtx = \snrrx$, and use the same design parameters as before.
% Note that, from \figref{fig:batman}, $\inr = 60$ dB was a point of disparity between our codebooks versus the conventional ones.
At low $\snrtx = \snrrx$, we notice that CBF offers higher sum spectral efficiency than CBF+Tay-40, thanks to its higher beamforming gain, which is more important at low \gls{snr} than interference suppression.
As $\snrtx = \snrrx$ increases, the interference mitigation offered by CBF+Tay-40 nets it a higher sum spectral efficiency over CBF.
% Note that CBF+Tay-20 begins to overtake CBF right around $\snrtx = \snrrx = 10$ dB, delayed versus its CBF+Tay-40 sibling due to its lessened self-interference suppression.
Our design, shown for various $\bitsphase = \bitsamp$, outperforms all designs across $\snrtx = \snrrx$ thanks to its robustness to self-interference even at this high $\inr$.
Notice that our design, along with CBF+Tay-40, reaches a high-\gls{snr} (approximately linear) regime much sooner than CBF and CBF+Tay-20 due to better self-interference rejection.
With higher $\bitsphase = \bitsamp$, as highlighted before, our design can achieve significant jumps in sum spectral efficiency, albeit with diminishing returns.

% \input{sec-simulation-results-v3.tex}

% \input{sec-simulation-results-v2.tex}

% \input{sec-simulation-results.tex}

% \input{sec-simulation-results-old.tex}

% \pagebreak

\section{Conclusion} \label{sec:conclusion}

To our knowledge, we have presented the first design of analog beamforming codebooks for \mmwave full-duplex, where we construct sets of transmit and receive beams that offer high beamforming gain while simultaneously reducing the amount of self-interference coupled between transmit-receive beam pairs.
% Our design accounts for digitally-controlled phase shifters and attenuators comprising practical analog beamforming networks and delivers guarantees on the beamforming gain supplied to users.
Numerical results show that our design outperforms off-the-shelf codebooks and confirms that full-duplex \mmwave systems can benefit from dedicated codebook designs offering robustness to self-interference without comprising beamforming gain.
With our codebooks, full-duplex \mmwave systems can support codebook-based beam alignment while also minimizing self-interference.
Our codebooks have the potential to improve a variety of existing \mmwave full-duplex solutions and can be supplemented by analog and digital self-interference cancellation.
Important future work includes characterizing the self-interference channel and designing codebooks robust to imperfect channel knowledge.

% \footnote{I.~P.~Roberts is supported by the National Science Foundation under Grant No.~DGE-1610403. Any opinions, findings, and conclusions or recommendations expressed in this material are those of the authors and do not necessarily reflect the views of the National Science Foundation.}

% It is important to keep in mind that all of these codebooks could be supplemented with additional self-interference cancellation measures to shrink the capacity gap, and our design demands significantly less of such.

% \input{sec-appendix.tex}

% \pagebreak

\section*{Acknowledgments}

% I.~P.~Roberts is supported by the National Science Foundation Graduate Research Fellowship Program under Grant No.~DGE-1610403. Any opinions, findings, and conclusions or recommendations expressed in this material are those of the author(s) and do not necessarily reflect the views of the National Science Foundation.

I.~P.~Roberts is supported by the National Science Foundation under Grant No.~DGE-1610403. 
Any opinions, findings, and conclusions or recommendations expressed in this material are those of the authors and do not necessarily reflect the views of the National Science Foundation.

% \pagebreak

% \section*{References} \label{sec:bibliography}
% \printbibliography[heading=none]
\bibliographystyle{bibtex/IEEEtran}
\bibliography{bibtex/IEEEabrv,refs}

\end{document}